\documentclass[acmsmall]{acmart}

\usepackage{booktabs}
\usepackage{tabularx}
\usepackage{multirow}
\usepackage{tablefootnote}
\usepackage{threeparttable}
\usepackage{pifont}
\usepackage[T1]{fontenc}
\usepackage{pbsi}
\usepackage{enumitem}
\usepackage[skins]{tcolorbox}
\usepackage[linesnumbered,ruled,vlined,noend]{algorithm2e}
\usepackage{caption}
\usepackage{listings}
\usepackage{subcaption}
\usepackage{xurl}
\usepackage{enumitem}
\usepackage{titlesec}

\newcommand{\TODO}[1]{{\textbf{\color{red}[[TODO: #1]]}}}
\newcommand{\Code}[1]{\begin{small}\fontsize{9.5}{10}\selectfont\texttt{#1}\end{small}}
\newcommand\revision[1]{{#1}}

\titlespacing*{\section}
{0pt}{0.8mm plus .4mm minus .4mm}{0.8mm plus .4mm minus .4mm}
\titlespacing*{\subsection}
{0pt}{0.8mm plus .4mm minus .4mm}{0.8mm plus .4mm minus .4mm}
\titlespacing*{\subsubsection}
{0pt}{0.8mm plus .4mm minus .4mm}{0.8mm plus .4mm minus .4mm}

\makeatletter
\patchcmd\algocf@Vline{\vrule}{\vrule \kern-0.4pt}{}{}
\patchcmd\algocf@Vsline{\vrule}{\vrule \kern-0.4pt}{}{}
\makeatother

\SetCommentSty{mycommfont}
\SetKwComment{Comment}{\ \#\ }{}
\SetKwProg{Func}{function}{}{end}

\tcbset{
  my box2/.style={
    enhanced,
    colframe=#1!80,
    colback=#1!5,
  },
}
\newtcolorbox{summary-rq}{
  my box2=black,
  boxrule=1pt,top=3pt,bottom=3pt,left=4pt,right=4pt
}

\AtBeginDocument{%
  }

\setcopyright{acmlicensed}

\setcopyright{acmlicensed}
\acmDOI{10.1145/3729345}
\acmYear{2025}
\acmJournal{PACMSE}
\acmVolume{2}
\acmNumber{FSE}
\acmArticle{FSE075}
\acmMonth{7}
\received{2025-02-25}
\received[accepted]{2025-04-01}

\acmConference[FSE 2025]{International Conference on the Foundations of Software Engineering}{Mon 23 - Fri 27 June 2025}{Trondheim, Norway}
\acmISBN{}  

\newcommand\toolname{{SmartNote}}

\begin{document}

\title{SmartNote: An LLM-Powered, Personalised Release Note Generator That Just Works}

\author{Farbod Daneshyan}
\orcid{0009-0005-6254-6273}
\authornote{Both authors contributed equally to this paper.}
\affiliation{%
  \institution{Peking University}
  \city{Beijing}
  \country{China}
}
\email{fabs@stu.pku.edu.cn}

\author{Runzhi He}
\orcid{0000-0002-6181-6519}
\authornotemark[1]
\affiliation{%
  \institution{Peking University}
  \city{Beijing}
  \country{China}
}
\email{rzhe@pku.edu.cn}

\author{Jianyu Wu}
\orcid{0000-0002-1785-215X}
\affiliation{%
  \institution{Peking University}
  \city{Beijing}
  \country{China}
}
\email{jianyu.wu@pku.edu.cn}

\author{Minghui Zhou}
\orcid{0000-0001-6324-3964}
\authornote{Corresponding author.}
\affiliation{%
  \institution{Peking University}
  \city{Beijing}
  \country{China}
}
\email{zhmh@pku.edu.cn}


\begin{abstract}
The release note is a crucial document outlining changes in new software versions. It plays a key role in helping stakeholders recognise important changes and understand the implications behind them. Despite this fact, many developers view the process of writing software release notes as a tedious and dreadful task. Consequently, numerous tools (e.g., DeepRelease and Conventional Changelog) have been developed by researchers and practitioners to automate the generation of software release notes. However, these tools fail to consider project domain and target audience for personalisation, limiting their relevance and conciseness.
Additionally, they suffer from limited applicability, often necessitating significant workflow adjustments and adoption efforts, hindering practical use and stressing developers.
Despite recent advancements in natural language processing and the proven capabilities of large language models (LLMs) in various code and text-related tasks, there are no existing studies investigating 
the integration and utilisation of LLMs in automated release note generation. 
Therefore, we propose \textsc{\toolname}, a novel and widely applicable 
release note generation approach that produces high-quality, contextually personalised release notes \revision{by leveraging LLM capabilities to} aggregate, describe, and summarise changes \revision{based on} code, commit, and pull request details. 
It categorises and scores (for significance) commits to generate structured and concise release notes of prioritised changes.
\revision{We conduct} human and automatic 
evaluations that reveal \textsc{\toolname} outperforms or achieves comparable performance to DeepRelease (state-of-the-art), Conventional Changelog (off-the-shelf), and the projects' original release note across four quality metrics: completeness, clarity, conciseness, and organisation. 
In both evaluations, \textsc{\toolname} ranked first for completeness and organisation, while clarity ranked first 
in the human evaluation. 
\revision{Furthermore, our controlled study reveals the significance of contextual awareness, while our applicability analysis confirms \textsc{\toolname}'s effectiveness across diverse projects.}

\end{abstract}



\maketitle

\section{Introduction}

Accompanying each new software version is a crucial document, the release note. Release notes summarise the new features, enhancements, bug fixes, and other changes in a software update, conveying the rationale and impact of changes to downstream developers and end users \cite{saner23, moreno2017arena}, effectively serving as a software trail \cite{abebe2016empirical}. 
To make informed update decisions, stakeholders 
evaluate the benefits of the update, such as bug fixes, features, and performance advancements, against potential drawbacks, like breaking changes that hinder the adoption of the new release \cite{Converge23:online}. 
Additionally, project managers 
use release notes to track the development progress and set milestones for release targets~\cite{bi2020empirical}. 

\begin{figure*}[ht]
    \centering
    \vspace{-1mm}
    \includegraphics[width=0.9\linewidth]{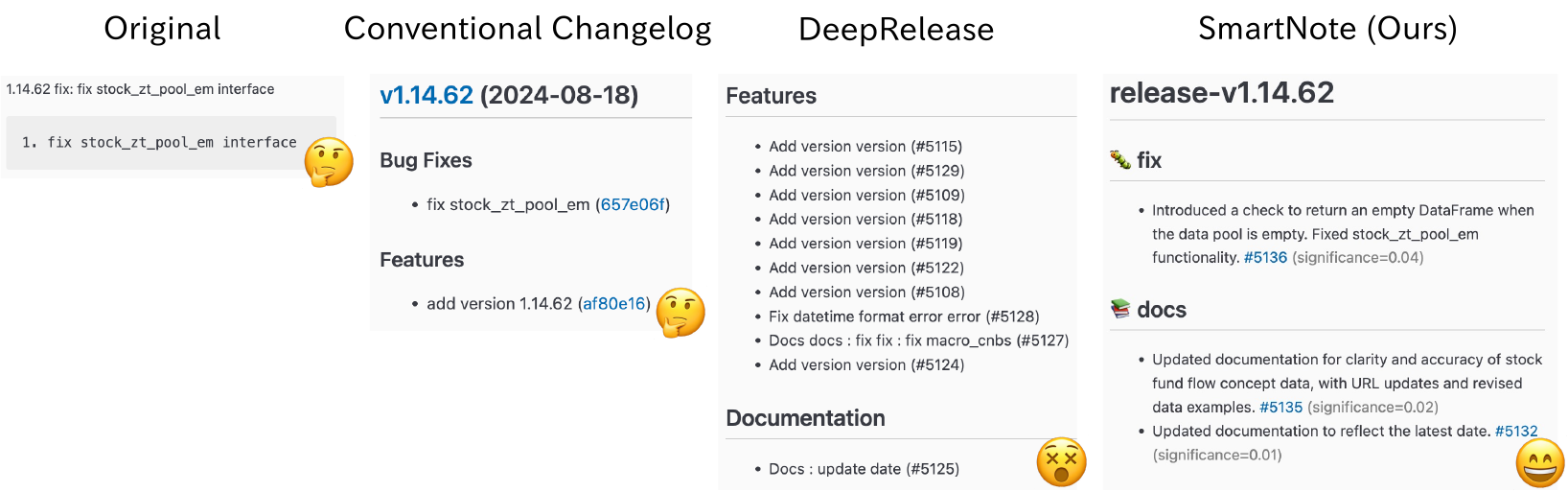}
    \vspace{-1mm} 
    \caption{Release Notes Generated for Version 1.14.62 of the AKShare Project \shortcite{akshare_example}.}
    \label{fig:teaser_image}
    \vspace{-1mm}
\end{figure*}

However, the task of writing release notes requires careful deliberation. Many developers and maintainers are reluctant as they perceive writing release notes as time-consuming and tedious; therefore, release notes are often neglected \cite{khomh2015understanding}. Developers have expressed their frustrations, with one Microsoft developer stating in a blog post \cite{MicrosoftHowToAutomateRN}, \textit{``We hate creating release notes when it's our turn to ship software updates. This can be quite a challenge and very time-consuming''}. Empirical evidence supports this claim. Moreno et al. \shortcite{moreno2014automatic} found that it takes up to 8 hours for an experienced developer to draft a release note document. This challenge is even more pronounced for the maintainers of large-scale open-source projects. For example, maintainers of \textit{espressif/esp-idf} are required to gather information from more than 5,000 commits to compile a single release note document for \textit{v5.2}. 
With the rise of continuous software delivery --- a software engineering approach that emphasises rapid, efficient, and reliable service improvements \cite{chen2015continuous, humble2010continuous} --- frequent software releases are becoming the new norm. This shift places an increasing burden on developers to consistently produce high-quality release notes. 
In general, release note producers face the time-consuming and challenging task of creating a well-balanced, high-quality release note,
while ensuring that it is suitable for their target audience \cite{2022ICPC-ReleaseNote}.

Automated release note generation tools such as \textbf{DeepRelease} \cite{jiang2021deeprelease} and \textbf{Convetional Changelog} \shortcite{conventional-changelog:online}, have been proposed and implemented by researchers and practitioners. 
However, they lack widespread adoption. 
Sumana et al. \shortcite{nath2024practitioners}, interviewed respondents who indicated that \textit{automated tools are not really useful to reduce work stress}. Similarly, Bi et al. \shortcite{bi2020empirical} found that none of the participants they interviewed had adopted such tools for automatically generating release notes. 
The lack of adoption may be due to 
limitations that hinder their real-world application. \textbf{First}, they often fail to consider the diverse needs of various audiences and project domains, which are important contextual factors for effective \textbf{personalisation}. For example, release note users and producers have different expectations on releases of different types (i.e., major, minor, patch) from projects of different domains (e.g., applications, tools) \cite{bi2020empirical, 2022ICPC-ReleaseNote, klepper2016semi}. This mismatch results in release notes lacking critical details for specific user groups and leaves users frequently overwhelmed with generic and irrelevant information. For instance, users of Libraries \& Frameworks prefer more prevalent changes documented in release notes, while users of Software Tools prioritise information about performance and security improvements \cite{saner23, nath2022exploring}.
\textbf{Second}, existing tools have limited \textbf{applicability}, demand extensive adoption efforts, require significant workflow adjustments, and often produce overly verbose outputs. For instance, ARENA only supports specific programming languages \cite{moreno2017arena}, and DeepRelease excludes 46\% of projects that do not adhere to pull request (PR) practices \cite{jiang2021deeprelease}. Our results 
confirm
that DeepRelease failed for \textasciitilde10\% of the projects we analysed. Other tools rely on basic automation that aggregates changes directly from version control systems, such as Git commit messages. However, this approach fails to adequately communicate the impact of changes \cite{openstack2022rm} and generates overly verbose release notes which lead to disengagement and difficulties in pinpointing significant and relevant changes \cite{saner23, nath2024practitioners}. 
Moreover, a common frustration with many off-the-shelf release note generators, like Conventional Changelog \shortcite{conventional-changelog:online} and semantic-release \shortcite{semantic-release:online}, is that they enforce commit message conventions, such as conventional commits \shortcite{ConventionalCommits:online} or angular \shortcite{angularj59:online} respectively, and require rigorous configuration. These approaches restrict usability by imposing requirements and pressuring developers to modify their workflows or design independent solutions. For instance, our evaluation reveals that Conventional Changelog fails for more than half the projects analysed.
\textbf{Third}, current approaches do not explore the \textbf{effectiveness of LLMs} in release note generation.
To the best of our knowledge, despite recent advancements in natural language processing (NLP) and the proven capabilities of LLMs in various code and text-related tasks \cite{achiam2023gpt, zhao2023surveylargelanguagemodels, touvron2023llamaopenefficientfoundation, claude2024sonnet35}, there are no existing studies investigating integration and utilisation of LLMs in automated release note generation. LLMs have significant benefits in code comprehension tasks that allow them to capture semantic meaning and connections between code and natural language text \cite{name2024using, feng2020codebert, wang2021codet5}, allowing for intricate summaries of changes. Therefore, the use of LLMs could offer benefits in aggregating information from various sources and enabling rich, high-quality, personalised release notes by automating the interpretation of complex commit histories, extracting the most relevant information, and incorporating writing styles and organisational patterns to automatically generate concise and accurate release notes. 


\revision{Thus}, to bridge the gap in existing literature and support open source software (OSS) maintainers, we propose \textsc{\toolname}\revision{, a} release note generator \revision{that utilises the highly effective code to natural language comprehension capabilities of LLMs \cite{name2024using} to produce} 
personalised, high-quality release notes for \revision{diverse} GitHub projects.
\revision{Notably, \textsc{\toolname} infers optimal settings based on project context and captures the semantics of code, commit messages, and pull requests. This enables it to generate comprehensive and contextually relevant release notes, as illustrated in Figure~\ref{fig:teaser_image}, even for projects where commit messages are inconsistent, unstructured, or non-compliant with conventional standards, enhancing its \textbf{applicability}.}
\revision{Moreover, \textsc{\toolname} is designed to accommodate the preferences of various project domains and release types \cite{saner23}, ensuring that generated release notes are contextually tailored in terms of content, organisation, and style.}
At a high level, \textsc{\toolname} \revision{achieves} this in four steps: 1) \revision{context comprehension,} 2) \revision{change summarisation,} 3) \revision{change categorisation, and} 4) \revision{change filtration to remove} less significant entries and details.
These steps are \revision{carried out through} a five-stage generation pipeline, as we will explain later in Section~\ref{pipeline_overview}.
\revision{For this study, we selected OpenAI's state-of-the-art LLM, "gpt-4o", which ranks \#1 on the LMArena leaderboard \cite{lmarena}. To optimise the prompts, we conducted multiple iterations of trial-and-error, guided by best practices in prompt engineering proposed in previous studies \cite{ekin2023prompt, wei2024cotp} and by LLM vendors \cite{sanders2024openaicookbook}.}

To evaluate \textsc{\toolname}, we analyse generated release notes for completeness, clarity, conciseness, and organisation by conducting human and automated evaluations on 23 
open source projects against baselines --- DeepRelease, Conventional Changelog, and the projects' original release notes. To begin with, we find that \textsc{\toolname} is applicable to all evaluated projects, \revision{whereas} DeepRelease fails \revision{for} \textasciitilde10\% and Conventional Changelog \revision{for} \textasciitilde54\% \revision{of projects}.
Furthermore, the human evaluation \revision{indicates} that over 80\% of participants agree or strongly agree that \textsc{\toolname} achieves the best results for completeness, clarity, and organisation, while conciseness ranks second. In the automated evaluation, we found that \textsc{\toolname} \revision{achieves} 81\% commit coverage, \revision{ranking} first in organisation with an information entropy of 1.59. \revision{In contrast, conciseness ranks third, yielding mixed results. This can be attributed to the compromises made by release note authors, who often improve conciseness by reducing commit coverage}. 
\revision{Moreover, to assess the impact of context awareness on release note quality, we conducted an ablation study with both automatic and human evaluations, comparing \textsc{\toolname}'s generated release note with three variants. The results reveal that \textsc{\toolname} captures human-written nuances, highlighting the importance of prompt engineering and context comprehension key components to \textsc{\toolname}’s effectiveness, particularly in completeness and clarity.}

\revision{To summarise, despite industry and academic advances in automated release note generation, a gap remains between current methods and developer expectations. 
\textsc{\toolname} resolves the current challenges by introducing the following innovations:}

\revision{\textbf{1. Workflow-Agnostic Design}: Unlike prior tools that rely on rigid workflows (e.g., pull-merge strategies used by only 54\% of OSS projects \cite{jiang2021deeprelease} or commit conventions which are varied and not widely adopted), \textsc{\toolname} works out-of-the-box with broader applicability (e.g., of the projects we evaluated, DeepRelease fails for more than 10\% and Conventional-Changelog fails for more than 50\%).}

\revision{\textbf{2. User-Centric}: \textsc{\toolname} generates clear, complete, and well-organised release notes, making them user-friendly, actionable, and tailored to diverse audiences, addressing gaps that hinder the adoption of state-of-the-art tools.}

\revision{\textbf{3. Tailored Pipeline}: \textsc{\toolname} uses a tailored pipeline to aggregate, classify, score commits for significance, merge related changes, and organise release notes based on project domain as defined by previous research \cite{saner23} and context. This approach ensures personalised, concise, and structured release notes.}

\revision{\textbf{4. Prompt Engineering}: Research shows that prompt engineering improves LLM performance \cite{wei2024cotp, liu2024pretrain, schulhoff-etal-2023-ignore, schulhoff2024promptreportsystematicsurvey}. Without it, LLMs produce inconsistent, verbose, and sometimes nonsensical results, whereas with it, they are better personalised and more applicable (Section \ref{chap:evaluation}, EQ2).}

\section{Background}

In this section, we focus on\revision{,}
1) studies analysing release note practices, characteristics, and usage; 
2) state-of-the-art tools for automating release note generation and an overview of off-the-shelf release note generation tools;
and 3) a summary concluding the limitations of current approaches.
\subsection{Release Note Characteristics and Contents}
Initially, release notes served as data sources for broader studies on software maintenance and evolution \cite{yu2009mining, alali2008s, maalej2010can, shihab2013studying}. It has only been in recent years that researchers have turned their attention toward empirical studies that specifically examine release note practices and the development of automated generation techniques.

Existing studies aimed at analysing release notes have greatly enhanced collective understanding of the characteristics and usages of release notes. \revision{Releases} contain \revision{various} aspects of information that \revision{must} be organised into categories, \revision{to form} well-\revision{structured} release notes\revision{,} which positively impact software activities \cite{bi2020empirical}\revision{,} such as software evolution and continuous software delivery. However, release notes can be overwhelming and tedious to write at times, resulting in missed information or errors \cite{2022ICPC-ReleaseNote}. Additionally, research \cite{nath2022exploring} finds that release note content contains four main types of artefacts: issues (29\%), PRs (32\%), commits (19\%), and common vulnerability and exposure (CVE) issues (6\%). \revision{It also highlights that} users prefer addressed issues to be summarised or rephrased\revision{,} and that users' preference for release note contents differs based on their background, for example, bug fixes are essential for developers and testers. Additionally, several studies find that conciseness is an important factor \cite{aghajani2020software} and that most release notes list only between 6\% and 26\% of issues addressed \cite{abebe2016empirical}.

Expectations for release notes vary across software project domains (such as application software, system software, libraries and frameworks, and system tools) and release types (major, minor, patch) \cite{2022ICPC-ReleaseNote}. These factors influence the structure and content of release notes, which in turn affects their applicability 
\cite{nath2024practitioners}. To this extent, Wu et al. \shortcite{saner23} conducted research to analyse and identify how release notes are organised, in what way they are written, and the content of release notes concerning project domain and release types. The authors analysed 612 release notes from 233 GitHub projects. They found that 64.54\% of release notes organise changes into hierarchical structures, sorted by change type, affected module, change priority (the most common), or a combination of the three. 
Additionally, they identify three types of writing styles, ``expository'' (directly list the title or content of change-related commits/PRs/issues), ``descriptive'' (rephrase the content of change-related commits/PRs/issues to increase the readability and summarise the content of similar commits/PRs/issues), and ``persuasive'' (provide additional information to help developers understand the changes, such as the rationale behind the changes, the impact of the changes) which provide additional information and enhanced explanations specific to the project domain. Finally, they analyse the contents of release notes from different project domains, finding ``System Software'' and ``Libraries \& Frameworks'' projects more likely to record breaking changes, while ``Software Tools'' projects emphasise enhancements. 

\subsection{Automatic Release Note Generation}
Researchers have developed various approaches and tools to help achieve and enhance release note automation. Klepper et al. \shortcite{klepper2016semi} introduced a semi-automatic approach to generating audience-specific release notes that integrates with continuous delivery workflows. While this method offers flexibility and reduces the burden on release note producers, it assumes ideal conditions and lacks detailed implementation guidance, which may limit its practical applicability in diverse development environments. 
For example, it assumes that the data sources (e.g., issue trackers and version control systems) are well-structured and consistently used by the development team, which may not always be the case in real-world projects.

Moreno et al. \shortcite{moreno2017arena} introduced ARENA (Automatic RElease Notes generAtor), a tool to automatically generate release notes for Java projects. 
In comparison, Nath et al \shortcite{nath2021towards} used extractive text summarisation technique, TextRank, to automate release notes production.
Their approach functions without predetermined templates and is language-agnostic. Similarly, Jiang et al. \shortcite{jiang2021deeprelease}, proposed DeepRelease, a language-agnostic tool utilising deep learning techniques to generate release notes from the textual information contained in PRs. The tool concatenates the title, description, and commit messages of a preprocessed PR as the input, summarising them into a single change entry for the release note. They found that 54\% of the 900 open-source projects they analysed generate release notes based on pull requests. However, while it is language-agnostic, it requires developers to follow specific pull request development practices, to automatically generate release notes. This requirement excludes the remaining 46\% of projects that do not adhere to these practices, a significant amount. In contrast, Kamezawa et al. \shortcite{kamezawa2022rnsum} introduced a summarisation approach leveraging transformer-based sequence-to-sequence networks.
This approach generates labelled release notes by summarising commit messages, making it adaptable to various projects without specific constraints. However, it places a significant burden on developers to consistently write high-quality commit messages.

Developers have also created off-the-shelf solutions like \textit{Conventional Changelog} \shortcite{conventional-changelog:online}, which by default requires users to adhere to the conventional commit convention \cite{conventional-changelog:online}. While less intricate than those proposed in research studies, off-the-shelf solutions are widely adopted (e.g., \textit{Conventional Changelog} has more than one million downloads per week \cite{conventional-changelog-downloads:online}) due to their accessibility and ease of integration into development workflows. 
Additionally, they are customisable to cater to different projects' needs, however, this flexibility necessitates extensive configuration due to their reliance on templates and labels. 
Moreover, these tools typically generate changelogs that list all changes, including minor revisions, using the commit message title to describe the change, whereas release notes focus on summarising the most significant changes in a new release, ensuring the impact is understandable for the user regardless of context \cite{openstack2022rm}. Both serve the same fundamental purpose. For example, Release Drafter \shortcite{releasedrafter:online}, provides users with the flexibility to configure labels and headings for their commits and release notes. However, this flexibility requires developers follow a specific commit pattern. Other tools such as \textit{semantic-release} \shortcite{semantic-release:online}, \textit{changelogen} \shortcite{parsa2024changeloggen:online}, and \textit{github-changelog-generator} \shortcite{korolev2024ghclg:online} operate on similar principles, enforcing commit conventions or specific patterns to generate changelogs.


\subsection{Knowledge Gap} \label{bg_summary}
Despite the strong interest in automation, \revision{the} application \revision{of release note generators} in the open source world remains limited. Bi et al. \shortcite{bi2020empirical} found that none of the participants they interviewed had adopted release note automation tools, and Sumana et al. \shortcite{nath2024practitioners}'s participants complain about the stress of release note production even with the assistance of automation.
A major contributing factor to this phenomenon is the limitation of current automation approaches.
Existing tools lack key features, are prone to errors, and are often difficult to configure \cite{2022ICPC-ReleaseNote, aghajani2020software}. Additionally, existing automation techniques often generate standardised patterns, which can confuse users and lead to hesitancy in adoption due to concerns for relevance and accuracy \cite{nath2024practitioners}.

\revision{To summarise,} existing research on release notes has highlighted the importance of structured, well-organised content that caters to both release note producers and users. However, current tools while effective in aggregating information and categorising release notes, have the following limitations:

\begin{enumerate}
    \item fail to address key discrepancies between release note producers and users; 
    \item do not personalise the content, structure and style to meet the diverse needs of various software domains and release types;
    \item have stringent requirements, such as commit convention, templates, labels, or PR strategies;
    \item require extensive configuration; 
    \item some work only for certain programming languages. e.g., \textsc{ARENA}.
\end{enumerate}
These limitations often \revision{hinder} widespread adoption, \revision{resulting} in inconsistent quality and detail in generated release notes. In short, current generation methods lack personalisation and applicability.

\section{Approach}



\revision{To address the} gaps and limitations in automatic release note generation, we introduce \textsc{\toolname}, a novel approach \revision{that leverages the comprehension capabilities of LLMs} \cite{name2024using} to aggregate the "what" and "why" of changes in code, commits and pull requests, \revision{generating} high-quality, personalised release notes \revision{while remaining language and workflow agnostic}. \revision{To personalise release notes to individual} projects and release \revision{types}, \textsc{\toolname} incorporates best practices \revision{in} content, writing style, and organisation identified in previous research \cite{saner23}. \revision{In this section, we first introduce the key component of \textsc{\toolname}, the LLM module. We then provide an overview, followed by a detailed explanation of each stage of \textsc{\toolname}.}

\subsection{The LLM Module}\label{chap:prompt_enginnering}
\revision{The LLM module is integral to the generation pipeline,} encapsulating capabilities for the release note generation process. \revision{It is responsible for formatting} input \revision{by generating prompts based on predefined templates and contextual information, as well as parsing} the output. \revision{Additionally, it manages} the model's context size limit \revision{by limiting the token count for each commit diff. \revision{In cases where the limit is exceeded, \textsc{\toolname} provides warnings}. However, the token limit is sufficiently large, making occurrences rare in practice, except in scenarios such as the initial commit or when entire folders are moved or deleted. In our evaluation, the model's context size limit was exceeded in only 4 of the 23 projects, affecting no more than two commits. To ensure future compatibility, \textsc{\toolname} is designed to function with any LLM. For this study, we employed OpenAI's "gpt-4o" model, which was considered state-of-the-art at the time and ranked \#1 on LMArena \cite{lmarena}.} The following subsections \revision{detail} the prompt engineering techniques \revision{employed} and the approach used to identify optimal hyperparameters.



\subsubsection{Prompt Engineering.}

\begin{figure*}[ht]
    \vspace{-1mm}
    \centering
    \includegraphics[width=0.9\linewidth]{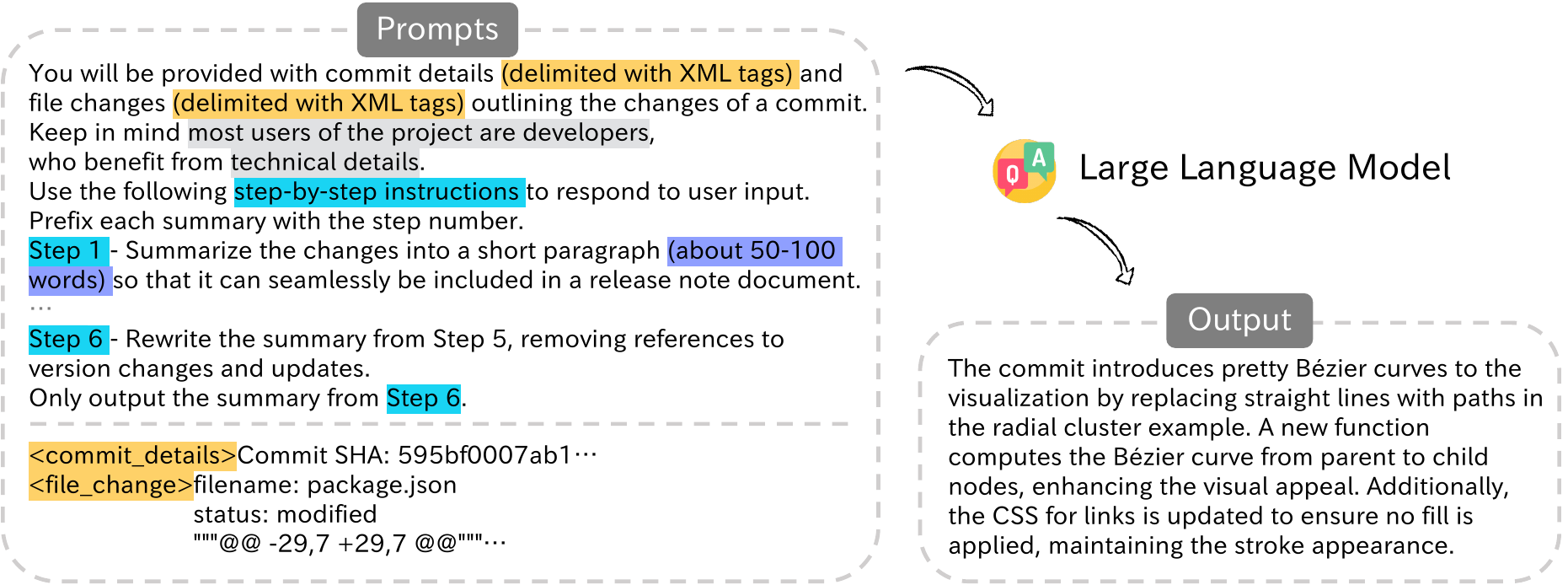}
    \vspace{-1mm}
    \caption{Model Input and Output Example for Commit Summarisation based on the d3 Project \shortcite{d3_prompt_example}.}
    \label{fig:commit_sum_model_in_out.png}
    \Description[An example of the model input and output.]{A diagram illustrating the model input and output for the commit summarisation task. Certain parts were removed to draw attention to the most important areas.}
    \vspace{-1mm}
\end{figure*}

To ensure high-quality responses from the LLM, we utilised prompt engineering techniques and best practices identified by research \cite{chen2024unleashing} and recommended by LLM vendors \cite{openai2024promptengineering}. Additionally, we iterated using a trial-and-error process to identify the most suitable prompt template --- manually analysing and evaluating the response, adjusting the prompt until the LLM outputs a correct and high-quality response.
Due to the page limit, we are unable to elaborate on all of the prompt templates. 
However, we illustrate the prompt engineering tactics with the example of the commit summarisation task in Figure \ref{fig:commit_sum_model_in_out.png} (which has been shortened to highlight the key aspects of the prompt and accommodate page space constraints). Complete versions of the prompt templates are available in our replication package.


\begin{itemize}
\item \textbf{Delimiters}: Used to clearly indicate distinct parts of an input, especially for multi-line strings, helping to correctly distinguish the data that the model should focus on with the instructions given. The more complex a task is the more important it is to disambiguate task details. We use XML tags \cite{openai2024promptengineering}, to clearly indicate the code patch when generating release note entries and when summarising PRs to clearly indicate details such as the title and body, and highlight the commits.
\item \textbf{Chain of Thought Prompting}: By providing a clear structure for the model to follow in the form of intermediate reasoning steps, the input helps guide the model's responses \cite{wei2024cotp, sahoo2024systematic}. To this extent, we split the complex task into simpler subtasks, and ask the model to follow a step by step guide to generate its response when summarising changesets.
\item \textbf{One-Shot \& Few-Shot Prompting}: Higher complexity tasks benefit from one-shot and few-shot prompting, a technique where one or a few examples are provided in the model input \cite{sahoo2024systematic}. While examples don't always help \cite{reynolds2021prompt, liu-etal-2020-multi}, they do serve as a mechanism to further guide the model in recalling previously learnt tasks. We used few-shot prompting in the settings generator to determine the project domain \cite{brown2020language} and one-shot prompting for rephrasing entries and content.
\item \textbf{Intent Classification}: Intent classification is a technique to identify the most relevant instruction for a query. In the case that the model needs to handle different cases, it is beneficial to categorise those cases ahead of time by hard coding it in the model input \cite{openai2024promptengineering}. We use this technique to specify the available project domains the model can use when identifying the project domain, ensuring the model outputs results we expect.
\item \textbf{Length Specification}: This technique involves specifying the length of the desired output \cite{openai2024promptengineering}. We use this during changeset summarisation to limit the model verbosity.
\item \textbf{Aligning the Decimals of Numbers}: 
The comprehension capability of LLMs is likely constrained by the design of BPE tokenisers, which interpret ``9.11'' as ``9'', ``.'', and ``11''. This limitation is still under discussion \cite{xie2024order}. Research reveals that this capability can be improved with number formatting \cite{schwartz2024numerologic}. Since adding special tokens to OpenAI's model is not viable, we used a simple yet effective technique to align the decimals to two digits.
\end{itemize}

\subsubsection{Hyperparameters.}
Hyperparameters affect the accuracy, verbosity, certainty and more of the LLMs' responses \cite{openai2024api}.
However, due to the high cost of inference, performing rigorous tuning (e.g., grid search) is not an option. Inspired by previous studies \cite{li2024onlydiff, liu2024your}, we greedy searched the following hyperparameters with manual evaluation:
\begin{itemize}
    \item \textbf{Temperature}: The temperature parameter controls randomness, with lower values producing more focused outputs \cite{openai2024api}. After experiments, we found a temperature of 0 yields consistent and satisfactory outputs, which aligns with previous studies \cite{peng2023makingchatgptmachinetranslation}.
    \item \textbf{Top-p}: Top-p implements nucleus sampling, considering only tokens within the specified probability mass. It controls the balance between certainty and creativity. We found a low top-p value of 0.1 encourages the LLM to generate deterministic output, and reduces the chances of generating casual content in few-shot classification tasks.
\end{itemize}

\subsection{\toolname\ Overview} \label{pipeline_overview}
\begin{figure*}[ht]
    \vspace{-1mm}
    \centering
    \includegraphics[width=1\linewidth]{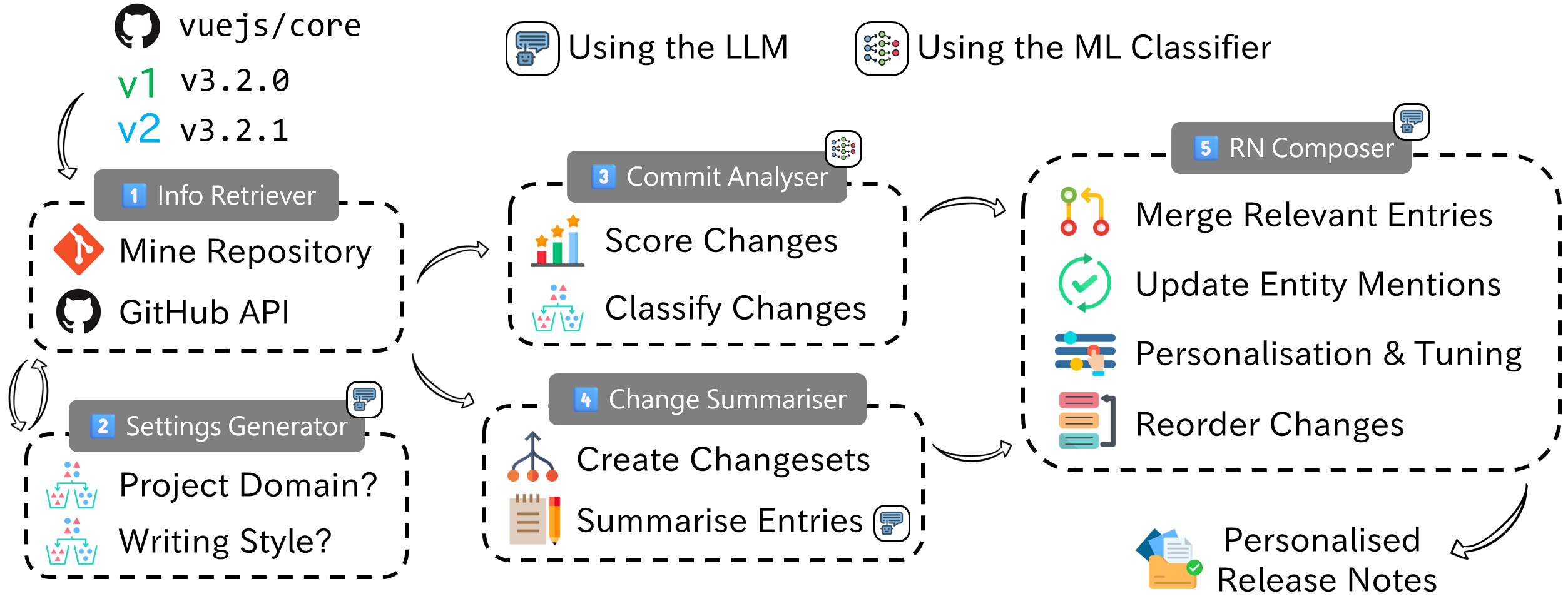}
    \vspace{-5mm}
    \caption{Overview of the \toolname\ Release Note Generation Pipeline}
    \label{fig:sd-dataflow-diagram}
    \Description[A data flow diagram illustrating the approach.]{A data flow diagram illustrating the different stages of \toolname. The approach is composed of 4 stages, (i) information retrieval, (ii) change summarisation, (iii) commit analysis, and (iv) formatting \& personalisation.}
    \vspace{-1mm}
\end{figure*}

In this section, we \revision{outline} the stages of \textsc{\toolname}\revision{'s generation pipeline} and \revision{describe its process for generating} a release note \revision{from} a GitHub repository \revision{given two versions, the previous version and the updated version}. As illustrated in Figure \ref{fig:sd-dataflow-diagram}, \textsc{\toolname} is composed of five stages: 
\revision{1) Info Retriever, 
2) Settings Generator, 
3) Commit Analyser, 
4) Change Summariser, and
5) RN Composer}.


In the first stage, information retrieval, \textsc{\toolname} collects all commit, pull request, and repository data associated \revision{with the changes made between the previous and updated versions.}
Next, \revision{in the} settings generation stage, \textsc{\toolname} starts by determining the project domain and commit message quality using the LLM module. 
Once all the data has been collected \revision{and the settings have been determined}, \textsc{\toolname} begins change summarisation. In this stage, commit data is packed into changesets which are then summarised into release note entries with the LLM module.
The commit analyser stage is run simultaneously, using the machine learning \revision{(ML)} classifiers to identify the conventional commit category and significance scores \revision{of commits}, prerequisites for filtering, summarising, and personalising the release note entries.
Finally, in the last stage, \textsc{\toolname} composes the release note. \revision{It} utilises the LLM module to refine \revision{the content} by first, rephrasing for conciseness and reorganising the entries, and then by \revision{adjusting} the release note to \revision{align with the} release context.

\subsection{\revision{Info Retriever}}

In the information retrieval stage, \textsc{\toolname} compares two versions of a GitHub project and collects data \revision{on} commits, pull requests, releases and projects.
First, \textsc{\toolname} looks for the corresponding git tags (a reference to a specific point in the repository's history \cite{git2024tagging}) of the previous release and the current release, and traverses over the commits with the PyDriller \shortcite{davide2024pydriller} library. In cases where this approach fails, \textsc{\toolname} falls back to the GitHub API. 
\revision{Notably,} we observed that Conventional Changelog fails when the release points to an orphan commit (e.g., \cite{questdb_off_tree_commit}), \textsc{\toolname} is not vulnerable in such cases.
Next, \textsc{\toolname} combines repository mining and the GitHub API to extract: 1)
release-related features such as release type (e.g., major, minor, patch), number of commits in the release, number of authors, and number of unique committers; 2) project-related features such as project name, previous versions and new versions, description, and README document; 3) commit-related features such as commit SHA, author, commit date, commit message, file patches (i.e., the code difference between the previous and new version or \textit{git diff}), file extensions, and lines changed; and 4) pull request features such as title, message, and associated commits. 
Additionally, the information retriever automatically identifies ``squash merge'' and ``rebase merge'' pull requests \cite{GitHubPRNamingPattern} and flags the associated commits --- a necessary step for merging individual commits into release note entries.

\subsection{\revision{Settings Generator}}
To \revision{accommodate the varying needs} of projects, releases, and target audiences, \textsc{\toolname} \revision{offers configuration options} for project domain, writing style, and structure, \revision{based on definitions from previous studies \cite{saner23}}. Additionally, it provides options for \revision{commit grouping and the minimum significance score (MST)}. \revision{However, configuring these options can be tedious. To simplify this, \textsc{\toolname} includes a settings generator that} is responsible for automatically identifying and \revision{applying context-aware defaults based on} findings from previous research and heuristics \revision{--- an approach proven effective in software automation tools \cite{he2023dependabot} --- making \textsc{\toolname} effectively zero-config.}

The \textbf{project domain} \revision{plays a key role in shaping} the writing and organisational style \revision{of the} release note \cite{saner23, moreno2014automatic}. \revision{\textsc{\toolname} does not} require users to \revision{interpret} domain definitions \revision{and choose} the \revision{most suitable domain}. \revision{Instead, it utilises} \revision{a} project domain classifier \revision{that leverages the LLM module to categorise the project domain based on the} project's description and README file. 

\revision{The \textbf{writing style} also plays a key role in shaping content and is closely linked to the project domain \cite{saner23}. If the user has not specified a writing style, \textsc{\toolname} automatically selects one based on the project domain. However, when the expository style is selected --- where release note entries are composed of the original commit message --- an additional evaluation step is taken to assess the project's suitability for this writing style. If it's unsuitable, the writing style is changed to persuasive. To achieve this, we designed a binary commit message classifier that leverages the LLM module. The prompts are engineered based on findings from a previous study identifying the characteristics of good commit messages \cite{tian2022commitmsg}. 
The classifier aligns closely with human-labelled data, achieving 95\% agreement (Cohen's kappa = 0.9) between two authors who, after carefully reviewing the definitions and examples of "good" commit messages outlined in a previous study \cite{tian2022commitmsg}, independently labelled the overall commit message quality for the projects and subsequently discussed their labels to reach consensus. 
}

\revision{The \textbf{MST} is an option that allows users to specify a minimum commit significance threshold for changes. It enables \textsc{\toolname} to filter out insignificant changes (e.g., fixing typos in comments or documentation; adjusting white space or indentation; and reformatting code without functional changes like running a linter), ensuring the release note is neither empty nor excessively long. This approach aligns with previous studies \cite{abebe2016empirical}, which found that well-formed release notes typically include only 6\% to 26\% of issues addressed in a release. It also reflects developer sentiments \cite{every_loses_readers}, which indicate that excessive detail can cause readers to lose attention, reinforcing the importance of conciseness.} \revision{Therefore, \textsc{\toolname}'s MST has been carefully tuned, ensuring release notes highlight significant changes without overwhelming the reader. This balance is achieved through iterative refinement and heuristic analysis, which determined that an MST between 0.1 and 0.15 strikes an optimal balance. Higher thresholds (0.2 or above) tend to introduce excessive detail, making release notes overly verbose, while lower (0.05 or below) thresholds risk omitting too much information, resulting in sparse release notes. For example, in cases like AkShare v1.14.62 \shortcite{akshare_example}, where most commits are minor or insignificant, increasing the MST helps include more changes. Conversely, in cases like RustDesk v1.3.0 \shortcite{rustdesk_example}, which has many meaningful commits, lowering the MST ensures a balance between verbosity and conciseness.}

\revision{The} \textbf{structure} \revision{determines how the release note is formatted. By default, it} is set to "Change Type", the \revision{most common} release note structure in open source \revision{projects} \cite{saner23}. \revision{Lastly} \textbf{commit grouping} \revision{determines whether commits are grouped together. It works by \revision{combining} commits associated with the same pull request into a single release note entry --- an approach commonly seen in real-world projects --- using the LLM module, thereby improving conciseness. This setting} is enabled by default\revision{,} as conciseness is critical.

\subsection{Commit Analyser} \label{chap:classifier}
\revision{
To address the problem of generic and verbose release notes, they are typically structured by change type \cite{saner23}, with included changes determined by their perceived importance. However, while developers are categorising changes in release notes, a vast majority of repositories do not categorise commits. Notably, approximately 90\% of the projects we sampled do not enforce a standard commit specification, explaining the limited effectiveness of off-the-shelf release note generators that rely solely on parsing commit messages. Moreover, projects varying in domain, scale, and collaboration methods may exhibit distinct development patterns and naming conventions. Accordingly, we design a commit analyser that accounts for variations in content preferences based on release type and project context \cite{bi2020empirical, saner23}. The commit analyser performs two tasks: 1) evaluating the relative significance of commits for conciseness and 2) categorising them for structural organisation. To achieve this, we leverage machine learning classifiers with models that are trained using contextual commit, release, and project features sampled from 3,728 repositories. To train the classifiers, we employ an approach that has been shown to be effective in software engineering studies \cite{xiao2022recommending, mariano2019feature}. This involves vectorising the text, combining it with numerical features, and feeding the resulting representation into XGBoost, an efficient, commonly used \cite{xiao2022recommending}, machine learning classifier. While most contextual features are numerical, LLMs have inherent limitations in processing numerical data effectively \cite{xie2024order}. In contrast, XGBoost, which implements the gradient boosting decision tree algorithm \cite{chen2016xgboost}, 
utilises a tree-based decision architecture, improving interpretability and ensuring transparency in the model’s decision-making process.
}

\subsubsection{Data Collection.}\label{chap:data_collection}
A high-quality dataset is important for creating reliable and precise models. We begin by defining our selection criteria: popular, actively maintained code repositories (i.e., excluding tutorials or resource collections) with a rich development and release history. To this end, we sample\revision{d} 3,728 non-forked and non-archived repositories on GitHub using the SEART GitHub Search Engine (seart-ghs). \revision{These repositories} were created before 2020, with more than 5000 stars, more than 10 releases on GitHub, and a codebase exceeding 100 lines. \revision{To ensure data quality, we removed six repositories whose git logs could not be parsed, as well as repositories that were duplicated or renamed. Additionally, we removed 23 repositories whose primary language was not English, as multilingual encoding introduces additional complexity and engineering overhead. Finally, we manually verified all repositories, removing five that contained tutorials or configuration files, resulting in a total of 644. The dataset was further processed for commit categorisation by retaining repositories in which at least two-thirds of the commits adhered to the conventional commits specification, resulting in 389 repositories. Lastly, to ensure the commit scoring model captures the intricacies of commit inclusion and exclusion in release notes, we refined the dataset to include only repositories which have more than three links to commits in their release notes, resulting in 272 repositories. From this dataset, we identified 21,882 releases encompassing 715,089 commits, of which 139,423 (17.7\%) were explicitly referenced in their corresponding release notes.}

\subsubsection{Feature Selection.} \label{feat_select}

To \revision{enhance} the accuracy and generalisability of the classifiers, we \revision{supply the model with} extensive contextual features:
\begin{itemize}
    \item \textbf{Commit Context:} The commit message \((emb*)\) is the most significant semantic feature, as high-quality commit messages conclude ``what'' and ``why'' of the changes \cite{tian2022commitmsg}.
    The scale of the changeset, i.e., the number of added lines \((\#addLines)\) and deleted lines \((\#delLines)\), represents the complexity of the commit and their relative importance \cite{levin2017boosting}. Furthermore, we build a programming language identifier using GitHub's linguist library \cite{github-linguist} to identify the programming languages of files modified in commits \((lang*)\), in order to support the commit classifier.
    \item \textbf{Release Context:} The type of release \((releaseType)\) influences the content of the release notes \cite{saner23}. We parse and compare release versions with the \textit{semver} library \cite{python-semver} into Major, Minor, Patch, and Unknown (version numbers incompatible with the semantic versioning specification). Moreover, we collect numerical contextual features representing the scale of the release: the number of commits between versions \((\#releaseCommits)\), the number of committers and the number of authors \((\#releaseAuthors)\). The modification complexity of the release is measured by averaging the number of modified files \((avgChangeset)\), lines of code affected per file \((avgCodechurn)\), and history complexity \((avgHistoryComplexity)\) \cite{hassan2009predicting}.
    \item \textbf{Project Context:} To quantify the project's scale and complexity, we measure the following metrics at the time of release: the number of commits \((\#commits)\), contributors \((\#contributors)\),  stars \((\#stars)\), issues \((\#issues)\), PRs \((\#prs)\), comments \((\#comments)\). We \revision{categorise} the projects into 4 domains \((projectDomain)\): \textit{Application Software}, \textit{System Software}, \textit{Libraries and Frameworks}, \textit{Software Tools} derived from the 6 project domains proposed by Borges et al. \shortcite{borges2016understanding} and studied by Wu et al. \shortcite{saner23}.
\end{itemize}
Commit messages were encoded into sentence embeddings (fixed-length vectors) leveraging the General Text Embeddings model (\Code{gte-base-en-v1.5}) \cite{li2023gtee}. With 137M parameters, it is small yet performant (ranking 31st on the MTEB classification leaderboard \cite{muennighoff2022mteb}), and ideal for \textsc{\toolname}'s intended use case --- CPU-only CI environments.
Statistics were mined with PyDriller, a Python framework that extracts information about commits, developers, modified files, diffs, and source code \cite{davide2024pydriller}. For historical events of the repositories, we utilised the GitHub API alongside the GHArchive, a project to record the public GitHub timeline, with the support of ClickHouse, a well-performing OLAP database.
Project domains were labelled manually by two authors. The first round of independent labelling yields a Cohen's kappa of 0.57, indicating moderate agreement \cite{mchugh2012interrater}; Two authors discussed and reached a consensus.

\begin{figure}[ht]
    \centering
    \begin{subfigure}{0.495\textwidth}
        \centering
        \includegraphics[width=\textwidth,trim={5 12 5 0},clip]{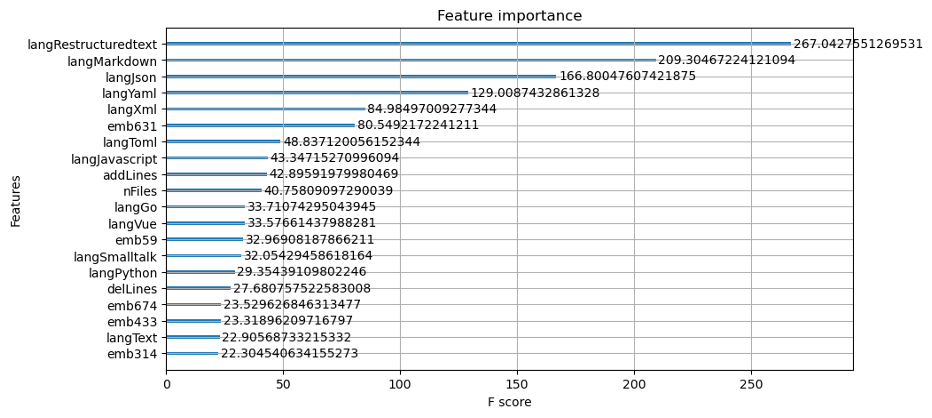}
        \caption{Commit Classification}
    \end{subfigure}
    \hfill
    \begin{subfigure}{0.495\textwidth}
        \centering
        \includegraphics[width=\textwidth,trim={5 12 5 0},clip]{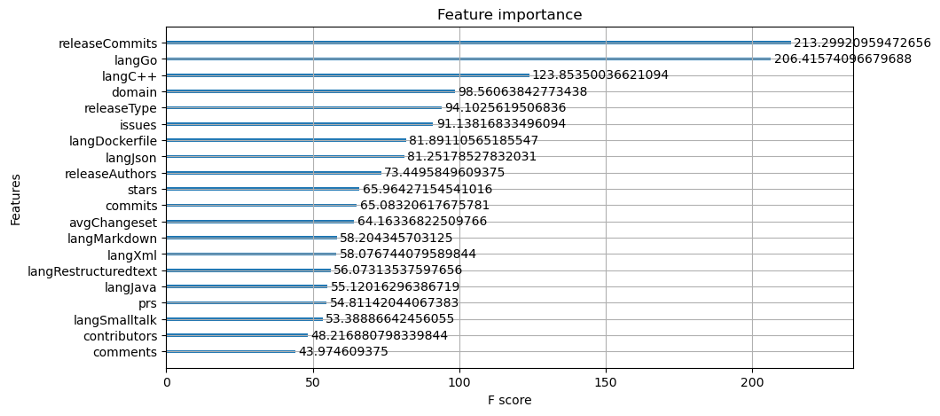}
        \caption{Commit Scoring}
    \end{subfigure}
    \vspace{-1.5em}
    \caption{Importance of Features in the Machine Learning Classifiers.}
    \label{fig:feature_imp}
    \vspace{-.5em}
\end{figure}

\subsubsection{Model Training and Feature Importance.}
\revision{
To mitigate the threat of overfitting, we implemented several strategies. First, we randomly partitioned the dataset into training, testing, and validation sets in a 7:2:1 ratio, following prior studies \cite{xiao2022recommending, jiang2021deeprelease}. Second, we applied the early stopping technique to terminate training when validation loss began to degrade. Third, we conducted K-Fold validation to ensure the randomness of the split. Finally, we determined the optimal hyperparameters using grid search.}


\revision{
The \textbf{commit classification model} converged after 100 iterations, with an accuracy of 0.71, a precision of 0.71, and a recall of 0.71 (numbers are weighted average). The 5-fold validation yielded the same performance numbers. \revision{Due} to the interpretability of decision trees, the importance of each feature is straightforward\revision{, as indicated by} XGBoost's reported gain value (the improvement in accuracy brought by a feature to the branches it is on) \cite{xgb_feature_imp} in Figure \ref{fig:feature_imp}. The programming languages of modified files (RST, Markdown, JSON, etc.) stand out, with several dimensions of the sentence embedding also being vital in the model's classification. The \textbf{commit scoring model} converged after 100 iterations, with a precision of 0.94 and a recall of 0.94. K-Fold validation yields consistent accuracy numbers between 0.94 and 0.95. Figure \ref{fig:feature_imp}, shows that the number of commits in the release is the most significant factor in the significance score, while domain, release type, and the number of issues also play key roles for determining commit significance. The project domain and release type are also notable factors, aligning with empirical evidence from previous studies \cite{saner23, bi2020empirical}.}

\subsection{\revision{Change Summariser}} \label{change_sum}
Gathering data from different sources is a common method that's utilised in existing research \cite{klepper2016semi} and ensures that the generated release note contains accurate and reliable information. In the change summarisation stage, \textsc{\toolname} combines data from the previous stages into changesets consisting of the commit date \& time, author, message, significance, change type, and file patches. 
The LLM module processes changesets, generating concise and accurate release note entries --- typically single sentences or brief paragraphs --- that form bullet points in the final release note.
By creating a changeset for each commit and summarising it individually, we break down the task into smaller, more manageable sizes for the LLM, preventing hallucinations, and resulting in higher quality summaries.
When commit grouping is enabled (which is by default), change summarisation combines all changes associated with a pull request into a changeset, summarises their descriptions, aggregates their scores, and replaces all associated release note entries with a single consolidated entry. 

\subsection{\revision{RN Composer}}
In the final stage, \textsc{\toolname} composes the release note by aggregating the changesets into a list using the identified organisation strategy, and then performs several smaller tasks \revision{to refine it}.

\begin{enumerate}
    \item \textbf{Merging Relevant Entries:}
    \revision{Occasionally, developers create multiple commits or pull requests for the same feature or change. This can be caused by bugs or alterations to the feature, causing overlapping entries that increase verbosity, decrease clarity, and cause reader confusion. With the exception of release notes with the "Change Priority" structure type, the LLM module is utilised to remedy this by merging related entries, consolidating them to improve clarity and readability, while ensuring no information is lost. E.g., in version 1.3.1 of Sniffnet \shortcite{sniffnetv131}, there were 22 commits referencing translation. In the release note produced by \textsc{\toolname}, these were consolidated into 10 entries.}
    \item \textbf{Updating \revision{Entity Mentions}:}
    In major refactoring releases, it's common for an entity \revision{(}a function or a variable\revision{)} to be added, renamed, or removed. 
    \revision{This causes fragmented entries that refer to the same entity as it worked or was named in different points of time in the past.}
    To address this inconsistency, 
    \revision{\textsc{\toolname} utilises the LLM module to identify renamed or modified entities and updates their mentions in the release note to match the most recent repository state at the specified version, improving coherence and clarity. E.g., in PR \#109 of the d3 project \shortcite{d3_prompt_example}, when a function was renamed based on feedback from the project maintainer.}
    \item \textbf{Personalisation and Tuning:}
    \revision{To improve conciseness, \textsc{\toolname} removes details based on the content preferences of the project and its audience. Specifically, entries with lower significance scores than the specified MST are removed; and the release note is trimmed and summarised using the LLM based on the project domain's description and content preferences}, guided by findings in \cite{saner23}. 
    \revision{E.g., AKShare v1.14.62 \shortcite{akshare_example}, which includes a single, simple code change. Without this step, a verbose 36-line release note is generated, with this step, changes are condensed into a single entry.}
    
    \item \textbf{Reordering Changes:}
    The inverted pyramid principle of writing suggests placing the most fundamental information at the top.
    Ordering with predefined rules may suffer from low generalisability though, as a previous study reveals that different projects prioritise different changes \cite{saner23}. \revision{\textsc{\toolname} utilises} the LLM to reorder the categories\revision{, moving breaking changes and new features, bug fixes, and enhancements to the top as they are considered more important \revision{\cite{abebe2016empirical, saner23}}. Document changes, dependency updates, and version changes are considered less important and placed lower}. \revision{E.g., in Bevy v0.14.1 \shortcite{bevy_no_prs}, features are moved to the top.}
\end{enumerate}

\section{Evaluation}\label{chap:evaluation}

\textsc{\toolname} is designed to utilise context-awareness to generate high-quality, personalised release notes whilst being broadly applicable. To this end, we design our evaluation to answer three questions: 1) Does \textsc{\toolname} generate high-quality release notes? --- necessary to understand in which aspects \textsc{\toolname} advances in; 2) Is \textsc{\toolname}'s personalisation method effective? --- to understand \textsc{\toolname}'s personalisation capabilities and effectiveness; and 3) How applicable is \textsc{\toolname}? --- to compare against existing tools and understand if there are any applicability limitations.

\begin{table}[ht]
\vspace{-2mm}
\caption{Distribution of project domains among the projects for evaluation.}
\vspace{-2mm}
\label{project-statistics}
\begin{tabular}{@{}llccl@{}}
\toprule
 & Project Domain & \multicolumn{1}{l}{Total Projects} & \multicolumn{1}{l}{\% of Projects} &  \\ \midrule
 & Application Software    & 7  & 30.43\% &  \\
 & Libraries \& Frameworks & 6  & 26.09\% &  \\
 & Software Tool           & 5  & 21.74\% &  \\
 & System Software         & 5  & 21.74\% &  \\
 & Total                   & 23 & 100\%   &  \\ \bottomrule
\end{tabular}
\vspace{-2mm}
\end{table}

To ensure a comprehensive comparison, we collected feedback for four types of release notes: first, for \textbf{\textsc{\toolname}}; second, for \textbf{\textit{DeepRelease}}, a state-of-the-art alternative, recent deep-learning power generator with accessible data; third, for the \textbf{original release notes} produced by the maintainers; and finally, for \textbf{\textit{Conventional Changelog}}, a remarkably popular off-the-shelf tool with more than one million downloads per week on NPM, a JavaScript package manager \cite{conventional-changelog-downloads:online}. 
This approach covers a wide variety of release notes, enabling us to effectively compare our work against existing methods and solutions. 


We began by brainstorming emerging open source projects \revision{from the  GitHub trending repositories page \shortcite{ghtrendingrepos}, considering programming languages such as Python, Rust, C++, C\#, Java, JavaScript, and TypeScript}.
We selected 33 projects with over 500 stars, under active development, and with recent community engagement 
(i.e., new commits, issues, discussions, or release history) and which allow for issues or discussions of enhancements, features or feedback. Another author independently reviewed the selected projects, ensuring they meet the criteria and identifying the project domain. 
The authors discussed any inconsistencies to reach a consensus. 
The projects were reevaluated \revision{for a} final time, 
and those that were historical, 
extremely large, 
seldom publish releases,
or appear to follow organisational guidelines to produce release notes
were excluded as they are unlikely to be interested in our study. 
Furthermore, we performed under-sampling 
to balance the representation of each domain. 
After reevaluation, 10 projects were removed, resulting in a total of 23 projects, as shown in Table \ref{project-statistics}. 
Next, we obtained their latest releases (7 minor releases and 16 patch releases) and release notes from the GitHub API.
Finally, we generate release notes using \textsc{\toolname}, DeepRelease, and Conventional Changelog. Note that
we used the default and automatic options of \textsc{\toolname}, to avoid overwhelming 
survey participants
with multiple variations of the release note. 
Additionally, we calculated the cost for \textsc{\toolname} to generate release notes for the evaluation projects with automatic settings. 
\textsc{\toolname}'s automatic release note generation costs an average of 90 cents per release (\$20.82 for 23 releases), which is economical given the extensive time typically required for high-quality release notes~\cite{moreno2014automatic}.

\subsection{EQ1: Does \textsc{\toolname} Generate High-Quality Release Notes?}

We conducted a human and an automatic evaluation to assess the effectiveness of \textsc{\toolname} and the release notes generated by it. 

\subsubsection{Quality Expectations.}

To measure release note content quality, we identify four key criteria from previous work:
\begin{enumerate}[parsep=-1pt]
    \item \textbf{Completeness}, to understand whether the release note sufficiently covers the changes made between the analysed versions. Missing or incorrect change information is a major issue in release note practices \cite{2022ICPC-ReleaseNote} and software documentation \cite{aghajani2020software}.
    \item \textbf{Clarity}, to understand whether the release note accurately and understandably reflects the changes made in the project. 
    The absence of clarity has been identified as a central issue in software documentation by previous studies~\cite{aghajani2020software}.
    The changes need to be explained with rationales and details to be understandable by target users.
    \item \textbf{Conciseness}, to understand whether the release note succinctly provides the right amount of information. 
    Studies confirm that conciseness is important \cite{aghajani2020software}. However, unnecessarily verbose release notes scare readers \cite{every_loses_readers} which explains why most release notes only list between 6\%-26\% of issues addressed in a release \cite{abebe2016empirical}.
    \item \textbf{Organisation}, to understand whether the release note is structured clearly and logically. 
    Practitioners prefer release notes to contain many different aspects of information and are organised into categories \cite{2022ICPC-ReleaseNote}. Well-formed release notes positively impact software evolution \cite{bi2020empirical}.
\end{enumerate}

\begin{table}[ht]
\centering
\vspace{-2mm}
\caption{\revision{Human and automatic evaluation results for release notes.}}
\vspace{-2mm}
\label{survey-results}
\begin{threeparttable}
\begin{tabular}{llccccl} 
\toprule
 & Criteria & \toolname  & DeepRelease & Original & \begin{tabular}[c]{@{}c@{}}Conventional \\ Changelog\end{tabular} &  \\ 
\midrule
 & Completeness & \textbf{4.00 (81\%)} & 3.39 (61\%) & 3.71 (74\%) & 2.74 (45\%) &  \\
 & Clarity & \textbf{4.06 (84\%)} &  2.97 (42\%) & 3.81 (77\%) & 2.71 (45\%) &  \\
 & Conciseness & 3.35 (55\%) &  3.03 (42\%) & \textbf{3.68 (65\%)} & 2.52 (29\%) &  \\
 & Organisation & \textbf{4.10 (84\%)} & 3.42 (55\%) & 3.52 (61\%) & 2.61 (35\%) &  \\
 \midrule
 & Commit Coverage & \textbf{81\%} & 41\% & 31\% & 13\% &  \\
 & Information Entropy & \textbf{1.59} & 1.04 & 0.78 & 0.45 &  \\
 & Token Count & 1162.48 & \textbf{251.70} & 498.52 & 2231.57 &  \\
 & Automated Readability Index & 33.06 & \textbf{26.19} & 30.93 & 102.38 &  \\
 & Dale-Chall Readability & \textbf{13.35} & 16.22 & 13.79 & 32.70 &  \\
 & Entity Percentage & 1\% & 7\% & 2\% & 4\% &  \\
 & Success Rate & \textbf{23 (100\%)} & 21 (90.48\%) & N/A & 15 (46.67\%) &  \\
\bottomrule
\end{tabular}
\vspace{-2mm}
\end{threeparttable}
\end{table}

\subsubsection{Human Evaluation.}


To conduct the human evaluation, we created a unique questionnaire for each project, allowing us to gather feedback from users most familiar with it and its domain. Using 5-point Likert scale questions, the questionnaire was designed to assess the four criteria identified earlier. 
Following this, we recruit participants with diverse backgrounds for the credibility of the survey.
First, we reached out to the most active maintainers of the projects via their public contact information. We received one response from a maintainer who was willing to participate in the survey. Next, we utilised snowball sampling to recruit ten open-source developers and six researchers with experience ranging from less than 1 year to over 8 years. In total, we recruited 17 participants, with the most common level of experience at 2 to 4 years, representing 55\% of participants followed by 4 to 8 years representing 29\% of participants. 
Participants selected up to 3 projects based on their familiarity, checked the projects' GitHub release pages, and completed questionnaires. The release note generator's name was masked to prevent bias.
Once a project received 3 responses, we removed it from the list, spreading the responses across all the projects. Table \ref{survey-results} presents the survey results, with scores averaged across all projects. 
The agreement percentage, shown in brackets, indicates the proportion of participants who either agreed or strongly agreed. \textsc{\toolname} outperforms all other tools in the categories of completeness, clarity, and organisation while conciseness achieves the second best performance.

\revision{The release notes for all evaluated projects are available in our replication package. As a glimpse, we highlight projects where \textsc{\toolname} rated significantly better, slightly better, or worse:}
\revision{
\begin{itemize}
    \item \textbf{Significantly Better}: AkShare (Figure \ref{fig:teaser_image}), LangChain, Sniffnet, and UniGetUI. In these projects, SmartNote produced release notes that were organised, context-aware, audience-specific, and prioritised significant changes while filtering out irrelevant details compared to raw LLM outputs and other tools. For example, the AKShare project where the commit messages and PRs are simple, \textsc{\toolname} addresses the changes understandably and concisely through code comprehension; in contrast, other tools borrow the developers' words and output nonsense. 
    \item \textbf{Slightly Better}: Zulip, StirlingPDF, es-toolkit, and Continue. These projects demonstrated improvements in clarity and organisation. SmartNote grouped related changes and provided well-structured summaries. Other tools and the raw LLM are likely to benefit from the fruitful commit messages and PRs in these projects and generate release notes of acceptable quality. While the improvements were less dramatic, the notes were still easier to interpret and more actionable than those generated by baseline tools.
    \item \textbf{Worse}: Jan is likely rated worse for the amount of dropped details (\textsc{\toolname}: 2 entries vs Conventional Changelog: 16 entries). The default MST configuration may have excluded changes that some users consider relevant. 
\end{itemize}}

The results of the survey reflect the performance of the fully automated pipeline. Considering this, \textsc{\toolname} outperforms both the original release notes and competing solutions in terms of completeness, clarity, and organisation. Over 80\% of participants agree that \textsc{\toolname} performs the best among the compared tools. While only 55\% agree regarding conciseness, this is unsurprising given our evaluation strategy. With additional project-specific tuning, users can achieve better results. In light of this, we can confidently say that \textsc{\toolname} is significantly superior to off-the-shelf tools and state-of-the-art alternatives. 

\subsubsection{Automatic Evaluation.}
Although human evaluation is undoubtedly the best way to assess release notes, we conducted an automatic evaluation to 1) complement our findings; and 2) explore the possibility of automatic quality assessment on RNs. 
Previous work \cite{jiang2021deeprelease, zhang2024automatic, li2024onlydiff} employed text similarity metrics such as BLEU and ROUGE for the purpose of comparing and measuring the differences between automatically generate\revision{d} release notes and gold standards.
However, we find their approach to \revision{unviable} for two reasons. First, previous studies that applied BLEU or ROUGE on generated release notes did not publish their evaluation dataset so it's impossible to compare. 
Second, text similarity metrics compare the lexical similarity of the text against ``gold references''; they can penalise semantically correct hypotheses if they differ lexically from the reference \cite{wieting-etal-2019-beyond}. \revision{Thus, they are likely} sub-optimal for the task\revision{, given} the length and diversity of high-quality release notes.



To address this, we designed six metrics to automatically measure the four quality aspects of our study. \textbf{First}, we used commit coverage to assess \textbf{completeness}, calculated by dividing the number of mentioned commits by the total number of commits in the release. A higher value indicates better overall completeness. \textbf{Second}, we measured \textbf{conciseness} using the token count of OpenAI's tokeniser, as the word count doesn't play well with code snippets. \textbf{Third}, to evaluate \textbf{organisation}, we leveraged markdown parsing to identify categories (headers) and items (bullet points), and we calculated information entropy, where a higher value signifies a higher number of categories and a more balanced distribution of entries, indicating better organisation. \textbf{Finally}, for \textbf{clarity}, we aimed to assess both specificity and understandability. We measured specificity by calculating the density of entities (specific software engineering terms, like the names of operating systems and libraries) with a \revision{specific software engineering} named entity recogniser \cite{nguyen2023software}, where a higher value suggests more technical details, and a neutral value is optimal. For understandability, we used the Automated Readability Index \cite{smith1967automated} calculated the average number of characters per word and of the words per sentence, where a lower score is better; and the Dale-Chall readability formula \cite{dale1948formula} to measure word commonality, where lower scores also indicate better readability.

Results in Table~\ref{survey-results} show that, compared to the next best alternative, \textsc{\toolname} achieves \revision{twice} the performance with 81\% coverage\revision{, whereas the} original release notes achieve only 31\% coverage. In terms of token count, which reflects conciseness, \textsc{\toolname} ranks third; however, this is not a concern, as users can easily adjust the level of conciseness through the configuration settings. These results suggest that release note authors tend to sacrifice coverage for conciseness, as indicated by the token count --- the number of words in the release note. In terms of organisation, \textsc{\toolname} ranks first, with an entropy of 1.59, a significant improvement over DeepRelease, the next best tool, which has an entropy of 1.04. This indicates that \textsc{\toolname} organises information much more effectively, consistent with the findings from our human evaluation. Finally, the automatic metrics for clarity yield mixed findings\revision{,} \textsc{\toolname} ranks third for the automated readability index but first for the Dale-Chall readability score, while having the lowest entity percentage, which measures specificity. 

Overall, \textsc{\toolname} \revision{demonstrates superior organisation and provides significantly better commit coverage compared to baseline methods}. These findings align with human evaluation\revision{s}, further \revision{confirming} that \textsc{\toolname} \revision{outperforms} off-the-shelf tools and state-of-the-art alternatives. 
\revision{In contrast,} \textsc{\toolname}'s rank for conciseness suggests that the default settings we used may not produce \revision{ideal} results. However, the results highlight that release note authors make sacrifices in order to compensate for other aspects\revision{, e.g., compromising on commit coverage to improve conciseness.} \revision{Therefore, reinforcing} the importance of giving users the control to make adjustments where necessary \revision{based on their} preferences. \revision{This demonstrates that \textsc{\toolname} is adaptable to both user and project needs, providing} a significant advantage over static, off-the-shelf solutions.


\vspace{-1.5mm}
\begin{summary-rq}
\textbf{Result of EQ1:} 
The human evaluation shows over 80\% of participants agree that \textsc{\toolname} performs best in completeness, clarity, and organisation. While only 55\% agree regarding conciseness. The automatic evaluation shows that \textsc{\toolname} ranked first in completeness with 81\% commit coverage and organisation with an information entropy score of 1.59. Conciseness ranked third with a token count of 1162.48 and clarity which showed mixed findings. Overall, the automated evaluation aligns with the human evaluation.
Conciseness did not achieve superior performance in either, but this is a non-issue as \textsc{\toolname} offers the flexibility to customise conciseness as preferred. 
\end{summary-rq}
\vspace{-1.5mm}

\subsection{EQ2: How Applicable is \textsc{\toolname}?}
To ensure applicability, it is important that release note generators can be used with all projects. Existing tools have stringent requirements, such as commit convention, templates, labels, or PR strategies; require extensive configuration; and some work only for certain programming languages. e.g., ARENA. \textsc{\toolname} addresses these limitations by\revision{:} \revision{1)} using a settings generator to determine optimal \revision{configurations}, \revision{2) ensuring}
developers do not have to follow any requirements, and \revision{3)} by being language-agnostic. \revision{Moreover, as shown in Figure \ref{fig:teaser_image}, \textbf{\toolname} adapts to the size and complexity of each project, producing concise and relevant release notes even for small projects.} \revision{To this extent,} we analyse the projects for which we generated release notes for to identify which succeeded and which failed (i.e., generating empty or completely wrong release notes). The success rate of each project is recorded in Table \ref{survey-results} \revision{and} shows that \textsc{\toolname} is widely applicable. Compared to existing tools, \textsc{\toolname} does not fail at generating release notes for any project, while DeepRelease and Conventional Changelog fail approximately 10\% and 54\% of the time respectively.
These failures can be attributed to the \revision{several main} limitations. In the case of DeepRelease, \revision{it is} only able to process PRs (e.g., bevy between version v0.14.0 and v0.14.1 \cite{bevy_no_prs}). \revision{While} in the case of Conventional Changelog, \revision{it is} unable to process off-tree commits (e.g., questdb \cite{questdb_off_tree_commit}), i.e., changes that are not between the two specified versions. \revision{Also, in cases where there's a lack of labelling and commit conventions, it does} not produce any output (e.g., flatbuffers between version v24.3.7 and v24.3.25 \cite{flatbuffers_no_output}).


\vspace{-1.5mm}
\begin{summary-rq}
\textbf{Result of EQ2:} 
\textsc{\toolname} successfully generates release notes for all projects, while DeepRelease fails for approximately 10\% of projects due to its stringent PR requirements and Conventional Changelog fails approximately 54\% of the time due its rigid commit convention requirement. 
\end{summary-rq}
\vspace{-1.5mm}

\begin{table}[b]
\vspace{-3mm}
\centering
\caption{\revision{Human and automatic evaluation results for the ablation study.}}
\vspace{-3mm}
\label{tab:ablated_survey_results}
\begin{tabular}{llccccl} 
\toprule
 & Criteria & Raw LLM & No Composer & Random Context & \toolname &  \\ 
\midrule
 & Completeness & 2.90 (41\%) & 3.21 (54\%) & 3.31 (55\%) & \textbf{4.00 (81\%)} &  \\
 & Clarity & 3.12 (57\%) & 3.31 (61\%) & 3.28 (55\%) & \textbf{4.06 (84\%)} &  \\
 & Conciseness & 2.91 (45\%) & 3.18 (53\%) & 3.11 (43\%) & \textbf{3.35 (55\%)} &  \\
 & Organisation & 3.04 (51\%) & 3.40 (63\%) & 3.87 (83\%) & \textbf{4.10 (84\%)} &  \\ 
\midrule
 & Commit Coverage & 30\% & 79\% & \textbf{82\%} & 81\% &  \\
 & Information Entropy & \textbf{2.04} & 1.21 & 1.42 & 1.59 &  \\
 & Token Count & \textbf{568.35} & 1029.22 & 970.17 & 1162.48 &  \\
 & Automated Readability Index & \textbf{14.46} & 33.22 & 30.44 & 33.06 &  \\
 & Dale-Chall Readability & \textbf{11.01} & 13.50 & 13.89 & 13.35 &  \\
 & Entity Percentage & 3\% & 1\% & 2\% & 1\% &  \\
\bottomrule
\end{tabular}
\vspace{-3mm}
\end{table}

\revision{\subsection{EQ3: How Effective is \toolname's Personalisation?}}
\revision{To better understand how \textsc{\toolname}'s contextually aware, personalised generation pipeline contributes to the quality of release notes, we conducted an ablation study with both automatic and human evaluations, comparing \textsc{\toolname}'s release note with three different variants:}
\vspace{-.01em}
\revision{\begin{itemize}
    \item \textbf{Raw LLM}: To understand the contribution of the LLM, we instruct it to generate a release note without any prompt engineering, guidelines or examples, relying solely on its world knowledge and comprehension capabilities.
    \item \textbf{No Composer}: A release note generated by \textsc{\toolname} without the RN Composer stage, a key component of its personalisation capabilities.
    \item \textbf{Random Context}: A release note generated by \textsc{\toolname} with a randomly selected project domain and an "Unknown" release type.
\end{itemize}}

\revision{For the human evaluation, we recruited a separate group of six participants, consisting of students and industry professionals, to rank release notes based on the four key aspects previously discussed: completeness, clarity, conciseness, and organisation.
To mitigate author bias between the previous survey and the ablation, we asked participants in both groups to score the generated release note for both the \textbf{Raw LLM} and the \textbf{Random Context} variant. Furthermore, to ensure a fair comparison, we applied weighted normalisation to our results, shown in Table \ref{tab:ablated_survey_results}.}

\revision{The results of the human evaluation for the \textbf{Raw LLM} variant show that participants consistently rated it lower across all metrics: completeness (4.00 vs. 2.90), clarity (4.06 vs. 3.12), conciseness (3.35 vs. 2.91), and organisation (4.10 vs. 3.04), confirming the significance of our prompt engineering efforts.
Further examination revealed:
1) Inconsistent Styles and Structures: E.g., a long, plain list of pull request titles and files changed after the organised list of changes for Manticore-Search v6.3.4 \shortcite{manticoresearch_example}.
2) Verbose Results for Simple Changes: E.g., a verbose 36-line release note for AKShare v1.14.62 \shortcite{akshare_example}, despite the change involving a single, simple code modification.
3) Minimal Technical Details and Practical Implications: E.g., "various code quality improvements and refactoring for better maintainability and readability" in a release note generated for QuestDB v8.1.0 \shortcite{questdb_v810_compare}. In comparison, the automatic evaluation yields mixed results. The commit coverage is lower and aligns with the human evaluation well, explaining why the token count is considerably lower when compared to other variants --- the LLM is sacrificing coverage for brevity. While initially, the information entropy indicates good organisation, we observe that an excessive information entropy score is not ideal in real world applications, due to its inconsistent categorisation and excessive granularity. E.g., in Zulip v9.1 \shortcite{zulip_v91_compare} where most commits would traditionally be categorised as documentation updates.}

\revision{Next, we examine the human evaluation for the \textbf{No Composer} and \textbf{Random Context} variants. As shown in Table \ref{tab:ablated_survey_results}, they similarly exhibit lower completeness and clarity, confirming that contextual understanding plays a significant role in generating more comprehensive and clearer release notes. Conciseness, however, presents a similar pattern to \textsc{\toolname}. While it's relatively lower across all variants, it still outperforms most baselines except for original, human written release notes (65\%), suggesting that release note authors tend to prioritise brevity by omitting details. These results indicate that the MST needs to be fine-tuned on a project-by-project basis to achieve balance between commit coverage and conciseness. 
On the other hand, organisation remains one of \textsc{\toolname}’s strongest aspects, with the No Composer variant, the worst performing one, still achieving 63\%, even without full contextual understanding. We attribute this to the LLMs world knowledge and code comprehension capabilities. In contrast, the automatic evaluation results indicate that the No Composer and Random Context variants perform comparably to \textsc{\toolname}. However, they do not align with the human evaluation, which indicates that while the release notes perform well in automated metrics, they fail to capture the nuances of human-written release notes, as seen in \textsc{\toolname}, which benefits from prompt engineering.}

\vspace{-1mm}
\begin{summary-rq}
\textbf{Result of EQ3:} 
\revision{To assess the impact of context awareness on release note quality, we conducted an ablation study with both automatic and human evaluations, comparing \textsc{\toolname}'s generated release note with three variants: Raw LLM (simply feed the changes to an LLM without any prompt engineering, guidelines or examples), No Composer (without the composition stage), and Random Context (random project domain).
The human evaluation of the Raw LLM, No Composer, and Random Context variants revealed that they are overly verbose and inconsistent. The automatic evaluation for the Raw LLM aligns with these finding. However, while the automatic evaluation of the No Composer and Random Context variants indicated metrics within margin of \textsc{\toolname}, they miss the nuance of human-written release notes, captured by \textsc{\toolname}, which is attributed to prompt engineering and context awareness. In summary, these findings highlight that context comprehension is key to \textsc{\toolname}’s effectiveness, particularly in completeness and clarity.}
\end{summary-rq}

\section{Limitations}
In this section, we discuss and address the limitations of our study, highlighting factors that may affect the validity of our study to guide future research. To this extent, we cover two factors: 
1) internal validity, and 2) external validity.

\subsection{Internal Validity} This concerns factors that are internal to the study which could potentially affect the study from accurately measuring what it intends to. 
Manually labelling the project domains in the dataset for training the classifiers may introduce author bias.
To mitigate this, two authors independently examined and labelled the data, and inconsistencies were discussed and resolved with a consensus.
Moreover, the variability of 
classifier accuracy across different projects (e.g., lower accuracy in projects with simple changes) could impact the results of the evaluation. For most projects in the evaluation set, the classifier's accuracy is approximately 0.6, which is acceptable given that a random classifier would achieve less than 0.1. 
However, for repositories where changes are simple and maintainers are not inclined to write meaningful release notes (e.g., AgentGPT), the classifier's accuracy can drop to around 0.3. 
To address this, we employ a range of different projects in our evaluation and recommend maintainers utilise commit message standards or prevent writing confusing and shorthand commit messages.
Automated commit message generators \cite{zhang2024automatic, li2024onlydiff} 
can be employed to enhance the quality of commit messages and release notes.
Additionally, the assessment of the release notes in our survey may be influenced by personal preferences and experience, which could affect the objectivity of the evaluation. This challenge is exacerbated by the absence of standardised evaluation criteria, potentially introducing bias into subjective judgements. To address this, a wide range of industry developers and PhD students were invited to participate in evaluations, thereby aiming to reduce bias through varied perspectives and enhance the overall objectivity and credibility of the assessment.

\subsection{External Validity} This concerns the generalisability of the findings of this study. First, \textsc{\toolname} has only been tested on OpenAI's ``gpt-4o'' model. However, the results of this paper should be generalisable to other top-tier LLMs (e.g., Claude, Gemini, Qwen) with another round of prompt engineering. We were not able to do so because of the high cost of LLM inference. Second, the human evaluation is performed on a relatively small evaluation set of 23 open-source projects, which may cast doubts on generalisability. However, the scale of evaluation is limited by the constraint of developer hours and communication efforts \revision{required} to conduct the survey. \revision{Additionally}, we \revision{ensured a diverse range} of projects and participants were represented: projects from varying domains, sizes, and popularity, and participants of various backgrounds were involved in the study.
Moreover, many off-the-shelf tools have features and quality-of-life aspects that maintainers may have become accustomed (e.g., first-time contributors). The absence of these features decreases the broad applicability of \textsc{\toolname}.


\section{Conclusion}
In response to the limitations of existing release note generators and off-the-shelf solutions, this study introduces \textsc{\toolname}, a novel, \revision{contextually tailored} automated release note generation approach designed for broad applicability and personalisation. \textsc{\toolname} leverages the latest in LLM technology and insights from empirical research to personalise to project domains and generate high-quality release notes without imposing stringent requirements or necessitating adjustments to existing workflows. 
In future research, we aim to implement and test \textsc{\toolname} in real-world scenarios to validate its effectiveness and contribute to the open-source community by building a tool with interactive editing capabilities that integrates into the release pipeline. 

\section{Data Availability}

The dataset and source code for this study are publicly available in a replication package on Figshare \shortcite{smartnote_repl_pack}.
For the complete codebase, setup instructions, and additional resources, please visit the GitHub repository: \url{https://github.com/osslab-pku/SmartNote}.

\section*{Acknowledgments}

This work is supported by the National Natural Science Foundation of China Grant 61825201 and 62142201.

\bibliographystyle{ACM-Reference-Format}
\bibliography{main}

\end{document}